\documentclass[a4paper,british,fleqn,usenatbib]{mnras}
\usepackage{ae,aecompl}
\usepackage[T1]{fontenc}
\usepackage[latin9]{inputenc}
\usepackage{array}
\usepackage{verbatim}
\usepackage{multirow}
\usepackage{amsmath}
\usepackage{amssymb}

\makeatletter

\providecommand{\tabularnewline}{\\}

\@ifundefined{date}{}{\date{}}
\usepackage{graphicx}

\newcommand{\jgrp}{J.~Geophys.~Res.~Planet.}
\newcommand{\cmda}{Cel.~Mech.~Dyn.~Astr.}
\newcommand{\mps}{Meteor.~Planet.~Sci.}
\newcommand{\advsr}{Adv.~Space.~Res.}

\title[%
Scattering V-type asteroids by jumping Jupiter
]{%
Scattering V-type asteroids during the giant planets instability:
A step for Jupiter, a leap for basalt.
}

\author[P. I. O. Brasil et al.]{
 P. I. O. Brasil,$^{1}$\thanks{E-mail: pedro\_brasil87@hotmail.com}
 F. Roig,$^{1}$
 D. Nesvorn{\'{y}}$^{2}$
 and V. Carruba$^{3}$
\\
$^{1}$ Observat{\'{o}}rio Nacional, Rio de Janeiro, RJ 20921-400, Brazil \\
$^{2}$ Southwest Reasearch Institute, Boulder, CO 80302, United States \\
$^{3}$ Universidade Estadual Paulista, Faculdade de Engenharia, Guaratinguet{\'{a}}, SP 12516-410, Brazil
 }

\date{Accepted XXX. Received YYY; in original form ZZZ}

\pubyear{2017}

\makeatother

\usepackage{babel}
\begin{document}
\label{firstpage}

\pagerange{\pageref{firstpage}--\pageref{lastpage}}

\maketitle
\begin{abstract}
V-type asteroids are a taxonomic class whose surface is associated
to a basaltic composition. The only known source of V-type asteroids
in the Main Asteroid Belt is (4)~Vesta, that is located in the inner
part of the belt. However, many V-type asteroids cannot be dynamically
linked to Vesta., in particular, those asteroids located in the middle
and outer parts of the Belt. Previous works have failed to find mechanisms
to transport V-type asteroids from the inner to the middle and outer
belt. In this work we propose a dynamical mechanism that could have
acted on primordial asteroid families. We consider a model of the
giant planets migration known as the jumping Jupiter model with five planets.
Our study is focused on the period of 
10~Myr that encompasses the instability phase of the giant planets. We show that,
for different hypothetical Vesta-like paleo-families in the inner
belt, the perturbations caused by the ice giant that is scattered
into the asteroid belt before being ejected from the solar system, are
able to scatter V-type asteroids to the middle and outer belt. Based
on the orbital distribution of V-type candidates identified from the
Sloan Digital Sky Survey and the VISTA Survey colours, we show that 
this mechanism is efficient enough provided that the hypothetical paleo-family 
originated from a 100 to 500~km crater excavated on the surface of (4)~Vesta.
This mechanism is able to explain the
currently observed V-type asteroids in the middle and outer belt,
with the exception of (1459)~Magnya. 
\end{abstract}
\begin{keywords} minor planets, asteroids: general \textendash{}
planets and satellites: dynamical evolution and stability\end{keywords}

\section{Introduction}

V-type asteroids are a particular taxonomic class of asteroids (e.g.
\citealp{2002Icar..158..146B}, \citealp{2009Icar..202..160D}) whose
surface mineralogy is associated to a basaltic composition (\citealp{2001M&PS...36..761B}).
Many V-type asteroids in the Main Belt, the so-called vestoids, belong
to a collisional family, the Vesta family (\citealp{1993Sci...260..186B};
\citealp{2014Icar..239...46M}), located in the inner belt ($2.1<a<2.5$~au).
This family originated by one or more cratering events that excavated
the basaltic surface of asteroid (4)~Vesta between 1 and 3~Gyr ago
(\citealp{2012Sci...336..694S}; \citealp{2012Sci...336..684R}; \citealp{2013JGRE..118.1991B}).
Many other V-type asteroids, that we will refer to as the non-vestoids,
do not belong to the current Vesta family, and cannot be directly
related to any specific collisional event. In particular, this is
the case of those V-type asteroids located in the middle ($2.5<a<2.8$~au)
and outer Main Belt ($2.8<a<3.2$~au), like (1459)~Magnya (e.g.
\citealp{2000Sci...288.2033L}; \citealp{2006Icar..183..411R}; \citealp{2008Icar..198...77M};
\citealp{2011epsc.conf..215D}). Neither large basaltic parent bodies
like (4)~Vesta, nor V-type asteroid families are known in these regions.
In this work, we investigate the possibility that some non-vestoid
V-type asteroids have been scattered from the inner belt during the
instability period related to the radial migration of the outer planets,
some 4~Gyr ago (\citealp{2005Natur.435..459T}; \citealp{2009A&A...507.1041M};
\citealp{2012AJ....144..117N}). 

Our focus is on the effect of the ``jumping Jupiter'' instability
of the giant planets (\citealp{2009A&A...507.1041M}) over the dynamics
of V-type asteroids originated in the inner belt. This instability
occurs while the giant planets are migrating by interaction with a
disk of planetesimals exterior to Neptune's orbit during the so-called
planetesimal driven migration (\citealp{1984Icar...58..109F}; \citealp{2005Natur.435..459T}).
The instability is related to a phase of mutual close encounters between
the planets, that eventually leads to the ejection of a Neptune size
ice giant after a scattering by Jupiter (\citealp{2011ApJ...742L..22N};
\citealp{2012AJ....144..117N}). In this context, the model requires
the initial existence of at least five planets: Jupiter, Saturn and
three ice giants. The encounters make Jupiter and the other giants
to undergo rapid and large variations of their orbits. In particular,
Jupiter's eccentricity is excited to its present value, and its semimajor
axis ``jumps'' inwards by $\sim0.3$\textendash 0.5~au. Also, the
period ratio between Jupiter and Saturn changes from an initial ratio
of 1.5 to the current $\sim2.5$ in a few tens of thousand years.
This behaviour satisfies the terrestrial planets constraint in that
Jupiter's orbit discontinuously evolves during planetary encounters
(\citealp{2009A&A...507.1053B}), thus avoiding secular resonances
which would otherwise lead to the disruption of the terrestrial planets
system. The model also satisfies several other constraints imposed
by the different populations of minor bodies (\citealp{2013ApJ...768...45N};
\citeyear{2014ApJ...784...22N}; \citeyear{2014AJ....148...52N};
\citealp{2014AJ....148...25D}; \citealp{2014AJ....148...56B}; \citealp{2015AJ....150...68N};
\citealp{2015AJ....150..186R}).

\citet{2016Icar...266..142B} have shown that the jumping
Jupiter instability with five giant planets can disperse beyond recognition
the asteroid collisional families that formed before or during the
instability, which would help to explain the current paucity of asteroid
families older than $\sim3$~Gyr (\citealp{2016MNRAS.tmp..322C}).
A key ingredient in this phenomenon is the scattering of asteroids
caused by the interactions with the fifth planet that is ejected of
the system. This ice giant can reach heliocentric distances as small
as 1.5\textendash 2.0~au before being ejected, thus sweeping the
asteroid belt during a short period of time. \citet{2015AJ....150..186R}
have shown that, in some cases, this scattering can lead the asteroids
to undergo semimajor axis changes of up to 0.5\textendash 1.0~au.
Therefore, the jumping Jupiter instability provides a possible mechanism
to implant V-type asteroids from the inner belt into the middle and
outer belt.

This paper is divided as follows: In Sect. 2, we discuss the distribution
of V-type and basaltic asteroids in the main belt and the previous
efforts to assess their dynamics and to understand their origin. Section
3 describes the methodology applied to this study. Our results are
presented in Sect. 4. The last section is devoted to conclusions.

\section{Basaltic asteroids in the Main Belt}

Basalt is expected to originate during the process of geochemical
differentiation in the interior of the largest asteroids. The paradigm
of this model is asteroid (4)~Vesta (\citealp{2012Sci...336..684R}),
a body of $\sim500$~km in diameter $D$, which has long ago identified
to own a basaltic crust (\citealp{1970Sci...168.1445M}; \citealp{2011SSRv..163...77Z}).
The differentiation process was probably more efficient in the inner
part of the Main Belt, where the temperature gradient favours the
condensation of refractory and volatile-poor elements, in particular,
the radioisotope $^{26}\mathrm{Al}$ whose decay is mainly responsible
for the heating in the asteroids interiors (\citealp{1993Sci...259..653G}).
This idea is supported by the fact that (4)~Vesta itself is in the
inner belt, and also by the fact that the fraction of volatile-rich
over volatile-poor asteroids with $D>50$~km is 3 to 8 times larger
in the middle/outer belt than in the inner belt (e.g. \citealp{2016A&A...588A..11M}).
The second largest known V-type asteroid is (1459)~Magnya, a $D\sim17$~km
body located in the outer belt (\citealp{2000Sci...288.2033L}; \citealp{2006Icar..181..618D}),
and the remaining known V-type asteroids, both vestoids and non-vestoids,
are all smaller than 7~km in diameter. In principle, it would not
be expected to occur differentiation in such small bodies, and similarly
to the vestoids and the Vesta family, the origin of non-vestoid V-types
should be most likely related to the collisional fragmentation of
large differentiated parent bodies (e.g. \citealp{2002Icar..158..343M};
\citealp{2007A&A...473..967C}).

\subsection{Dynamics}

So far, the only confirmed source of basaltic material in the Main
Belt is (4)~Vesta. Therefore, several works have been devoted to
try to establish a dynamical link between (4)~Vesta and the non-vestoid
V-type asteroids. The interplay between mean motion resonances, secular
resonances, and the Yarkovsky effect (\citealp{2002aste.conf..395B})
has been addressed by \citet{2005A&A...441..819C} and \citet{2008Icar..193...85N},
who found that many non-vestoids in the inner belt can originate as
fugitives from the Vesta family over Gyr timescales. These mechanisms,
however, are not enough to explain a set of V-type asteroids in the
inner belt showing lower orbital inclinations, on average, than the
current Vesta family. \citet{2007A&A...465..315C} studied the effect
of close encounters between small vestoids and (4)~Vesta, and concluded
that the role of this mechanism in generating non-vestoids is minor.
\citet{2008Icar..194..125R} and \citet{2014CeMDA.119....1F} analysed
the crossing probability of small vestoids through the 3:1 mean motion
resonance with Jupiter, driven by the Yarkovsky effect. They found
that this mechanism would not be enough to explain the presence of
several km-size V-type asteroids in the middle belt. Finally, \citet{2011JPhCS.285a2024R}
considered the transport of V-type asteroids from the inner belt through
a Mars crossing regime combined with resonance stickiness, and concluded
that this mechanism is not able to implant V-type asteroids beyond
2.5~au in long term stable orbits. In summary, long term dynamics
proved to be insufficient to link a significant fraction of the non-vestoid
V-types to (4)~Vesta. 

\begin{figure}
\begin{centering}
\includegraphics[width=0.73\columnwidth,angle=270]{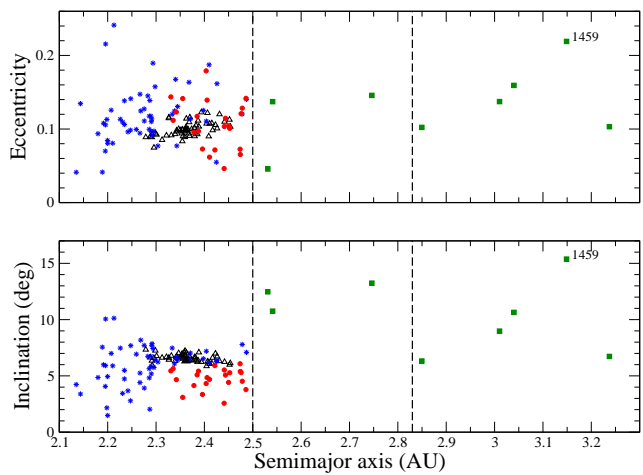}
\par\end{centering}
\caption{ The orbital distribution of the 138 spectroscopic V-type asteroids
in the Main Belt, compiled from different authors (see text). The black full triangle represents
(4)~Vesta. The different populations are represented, respectively, by black open triangles
(vestoids), blue stars (fugitives), red full circles (low inclination
inner belt), and green full squares (middle/outer belt).
(1459)~Magnya is indicated by its number. The vertical dashed lines
indicate the approximate boundaries of the inner, middle, and outer
asteroid belt.}
\label{Brasil-fig1}
\end{figure}

Following these results, we will refer hereafter to four
dynamical populations among the V-type asteroids:
\begin{itemize}
\item The vestoids, which correspond to the Vesta family members.
\item The fugitives population, i.e. the non-vestoids in the inner belt
that can be dynamically related to (4)~Vesta and/or to the Vesta
family; these correspond to objects with $a\lesssim 2.3$~au OR $I\gtrsim 6^{\circ}$.
\item The low inclination inner belt population, i.e. the non-vestoids 
in the inner belt that show no dynamical link to (4)~Vesta at all;
these correspond to objects with $a\gtrsim 2.3$~au AND $I\lesssim 6^{\circ}$.
\item The middle/outer belt population of non-vestoids, which
correspond to objects with $a>2.5$~au.
\end{itemize}
Figure \ref{Brasil-fig1} summarises the orbital distribution of the 138 currently
known V-type asteroids in the Main Belt confirmed from spectroscopic
observations, and identifies the different populations defined above. 
This figure compiles data from \citet{2009P&SS...57..229D},
\citet{2011epsc.conf..215D}, \citet{2011A&A...533A..77D},
\citet{2016MNRAS.455.2871I} and \citet{2017MNRAS.464.1718M}.

 In the above context, (1459)~Magnya and the other middle/outer belt
V-types, in particular, result to be the most paradoxical cases.
The possibility of a local origin of (1459)~Magnya was first addressed
by \citet{2002Icar..158..343M}, who proposed that, if this asteroid
formed from the collisional breakup of a larger differentiated parent
body in the outer belt, the subsequent family of Magnya would be dispersed
beyond recognition over hundreds of Myr timescales. However, even
if such hypothetical family would not be recognised nowadays as a
dynamical family (i.e. a cluster of asteroid sharing similar orbital
properties), footprints of its existence should be present in the
outer belt. In particular, we should expect to find other V-type asteroids
there besides (1459)~Magnya\footnote{In a different context, \protect\citet{2016MNRAS.tmp..322C} found
that asteroid paleo-families (i.e. families formed about or more than
4~Gyr ago) would be able to dynamically disperse beyond recognition
over the age of the Solar System, but footprints of them should be
still detectable in the Main Belt.}. As we can see in Fig. \ref{Brasil-fig1}, the shortage of V-type asteroids
beyond 2.5~au contrasts with the over abundance observed in the inner
belt, where the Vesta family formed. This shortage may partly be due
to observational bias, but a similar trend is observed when considering
candidate V-type asteroids selected from the visible colours of the Sloan
Digital Sky Survey (SDSS; 
\citealp{2006Icar..183..411R}; \citealp{2008Icar..198...77M};
\citealp{2010A&A...510A..43C}; \citealp{2014MNRAS.439.3168C}; \citealp{2014MNRAS.444.2985H}),
as well as from the infrared colours of the VISTA Survey
(MOVIS; \citealp{Licandro2017}). Actually, the fraction of V-type
asteroids in the middle/outer belt with respect to the total number
of V-type asteroids is of only $\sim4$-6\%
not only in the spectroscopic sample shown in Fig. \ref{Brasil-fig1},
but also in the SDSS (e.g. \citealp{2010A&A...510A..43C}) and the MOVIS
samples. This, together with the fact that
volatile-poor materials would not have been abundant in the outer
asteroid belt, seem to point against the idea of a local origin of
(1459)~Magnya.

Concerning the other V-type asteroids in the middle/outer belt, \citet{2007A&A...473..967C}
first proposed that some middle belt V-types could be fragments of
a collisional disrupted differentiated parent body whose best candidate
would be asteroid (15)~Eunomia (\citealp{2005Icar..175..452N}), located at 
2.64~au. 
Eunomia has associated a collisional family, and the three spectroscopic V-types 
presently known in the middle belt could be interpreted as dynamical fugitives from 
this family. On the other hand, no spectroscopic V-types are known within the Eunomia 
family; only some SDSS V-type candidates were actually reported as family 
members (\citealp{2014MNRAS.439.3168C}), but
none of these candidates have been observed spectroscopically so far.
\citet{2014MNRAS.439.3168C} and \citet{2014MNRAS.444.2985H} analysed the dynamics of the SDSS
photometric V-types in the middle/outer belt, and concluded that they
are segregated into eight regions in the $a,I$ plane that are dynamically
isolated from each other. This led the authors to propose that at
least eight different sources of basaltic material, either local or from Vesta, would 
be necessary to explain the current sample of candidate V-types beyond 2.5~au.

\subsection{Surface properties}

Observational evidence indicates that some V-type asteroids in the
Main Belt show mineralogical properties that are incompatible with
those of (4)~Vesta (e.g. \citealp{2004Icar..167..170H}; \citealp{2006AdSpR..38.1987D};
\citealp{2009P&SS...57..229D}), suggesting an origin from other (unknown)
basaltic parent bodies. Recently, \citet{2016MNRAS.455.2871I} gathered
the available spectroscopic observations of several known V-type asteroids
and analysed them using a uniform procedure to determine their mineralogical
properties. They found that (1459)~Magnya and the other non-vestoids
located beyond 2.5~au have spectroscopic properties somehow distinctive
from those of the vestoids and non-vestoids in the inner belt, supporting
the idea that (1459)~Magnya and the other middle/outer belt V-type
asteroids are not genetically related to (4)~Vesta.

\citet{Gil-Hutton2017} reported polarimetric measurements of 28 Main Belt V-type asteroids,
and found that they can be classified into two groups: (i) those that
show measurements compatible with the polarimetric curve of (4)~Vesta,
and (ii) those that show measurements compatible with the polarimetric curve 
of (1459)~Magnya%
\footnote{ \citet{2016MNRAS.456..248C} have shown 
that the polarimetric curve of (4)~Vesta
presents variations related to different albedo, composition and/or 
rugosity over its surface. Nevertheless, this variations are not enough to account
for the differences with respect to the polarimetric curve of (1459)~Magnya.
In particular, the inversion angle of Vesta is $\sim 2^{\circ}$ greater than
that of Magnya, and this paremeter is not sensitive to the changes on Vesta's 
surface.}. 
The first group includes several vestoids and inner belt non-vestoids,
while the second group includes some inner belt vestoids and non-vestoids, as well as 
middle/outer belt non-vestoids.

It is worth noting that the errors
involved in the determination of spectroscopic and polarimetric parameters
are usually large. In many cases, the differences observed among the
samples fall within their 1$\sigma$ uncertainties, making them indistinguishable in practice. 
In principle, there is no
clear relation between these spectroscopic and polarimetric differences
and the different dynamical populations.

\subsection{Size distribution}

So far, the clearest evidence that prevents a link between (4)~Vesta
and (1459)~Magnya is the size of the latter, which does not fit within
the expected size frequency distribution (SFD) of the ejecta from craterization
events (\citealp{2007Icar..186..498D}; \citealp{2015arXiv150203929M})
like those that excavated the surface of (4)~Vesta. On the other
hand, the remaining non-vestoids (from both the spectroscopic and
SDSS samples) are all as small as the vestoids, and may also have
originated from craterization events. Images of Vesta's surface taken
by the Dawn probe revealed two large basins in the south pole of the
asteroid. One of them, named Rheasilvia, with $\sim500$~km of diameter
has been dated $\sim1$~Gyr ago. The other one, named Veneneia, with $\sim400$~km of diameter has
been dated 2-3~Gyr ago (\citealp{2012Sci...336..690M}).
While Rheasilvia is the likely source of the current
vestoids (\citealp{2013JGRE..118..335M}; \citealp{2013JGRE..118.1991B}),
 Veneneia could be the source of several non-vestoids 
(\citealp{2012Sci...336..694S}). \citet{2016Icar..271..170P}
studied the formation of large basins on the surface of (4)~Vesta
during the epoch of the giant planets instability, and concluded that
any record of such basins would have been erased by the subsequent
crater saturation over the age of the solar system. This is in line
with the fact that observed craters on Vesta's surface dated less
than $\sim 4.1$~Gyr ago (\citealp{2014P&SS..103..131O}). 
Therefore, the possibility of a very early craterization event
on the surface of (4)~Vesta that led to the origin of a Vesta-like
family cannot be ruled out.

\begin{figure}
\begin{centering}
\includegraphics[width=1.0\columnwidth]{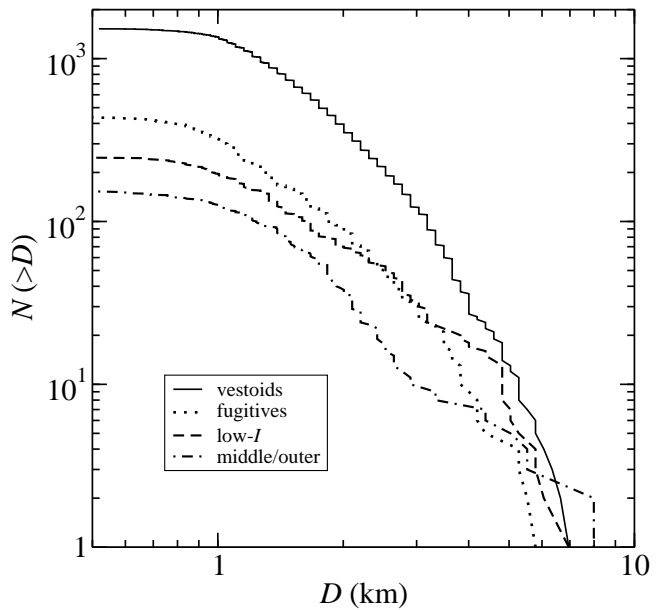}
\par\end{centering}
\caption{The cumulative size frequency distribution of the photometric V-type
candidates in the Main Belt observed by the SDSS. Diameters are estimated
assuming an albedo of 0.4. The different populations refer to those 
shown in Fig. \protect\ref{Brasil-fig1}.}
\label{Brasil-fig2}
\end{figure}

In this context, it is interesting to discuss the SFD
of the different dynamical populations of V-type asteroids.
This is shown in Fig. \ref{Brasil-fig2} for the sample of SDSS V-type
candidates, but a similar result was obtained by \citet{2008Icar..193...85N}
considering the spectroscopic sample. The fugitives population shows
a similar SFD to the vestoids, indicating that these two populations
may be related and have the same collisional age. On the other hand,
the low inclination population and the middle/outer population show
SFDs that differ from the Vesta family, suggesting that these populations
have a different origin/age. These populations (excluding (1459)~Magnya due to its 
size)
could have been originated by early craterization events on the surface
of (4)~Vesta, like the Veneneia basin or even older events 
(\citealp{2012Sci...336..694S}; \citealp{2013JGRE..118..335M}). 

\section{Methods}

Our goal in this study is to determine whether or not the jumping
Jupiter instability could explain the orbital distribution of the
non-vestoids, in particular, those in the middle/outer belt, assuming
that they formed on a very early craterization event on (4)~Vesta's
surface. We have proceeded in the same way as described by \citet{2016Icar...266..142B}.
In short, we have generated Vesta-like synthetic families, i.e. artificial
families with orbital parameters similar to those of (4)~Vesta, using
an algorithm that produces an initial orbital distribution of $N$
family members assuming that their ejection velocities $v_{\mathrm{ej}}$ from the parent
body follow a Maxwell distribution, with given mean $\bar{v}_{\mathrm{ej}}$.
The smaller the mean ejection velocity, the more compact the initial
family. Orbital elements of each member are then obtained from the
ejection velocities using Gauss formulas, for given values of the
orbital elements of the parent body: semimajor axis $a$, eccentricity
$e$, inclination $I$, true anomaly $f$, and argument of perihelion
$\omega$. In all cases, we have considered $a=2.36$~au. We have
also considered $f=90^{\circ}$ and $f+\omega=0^{\circ}$, which produces
rather spherical distributions in the space of orbital elements. 
The other parameters have been chosen in order to produce four different
orbital setups: (i) a low-$e$ and low-$I$ initial family ($e=0.089$, $I=7.14^{\circ}$),
(ii) a low-$e$ and high-$I$ family ($e=0.089$, $I=15^{\circ}$), (iii)
a high-$e$ and low-$I$ family ($e=0.2$, $I=7.14^{\circ}$), and (iv) a high-$e$
and high-$I$ family ($e=0.2$, $I=15^{\circ}$). In addition, for each of these
four setups, we considered two different values of $\bar{v}_{\mathrm{ej}}$:
100 and 500~m~$\mathrm{s}^{-1}$, representing a very compact and a very dispersed 
initial family, respectively. The initial number of members, $N$, was set to 1000 
for the compact families and 5000 for the dispersed ones.
This makes a total of eight different initial configurations.

The family members have been considered as massless test particles,
subject to the gravitational perturbation of the giant planets only.
To simulate the evolution of the giant planets during the jumping
Jupiter instability, we have used an hybrid version of the SWIFT\_RMVS3
symplectic integrator (\citealp{2015AJ....150..186R}), that reads
the positions and velocities of the planets from a file where they
were previously stored at 1~yr intervals, and interpolates them to
the desired time step using a two-body approach (\citealp{2011ApJ...742L..22N}).
The jumping Jupiter evolutions have been previously developed by \citet{2012AJ....144..117N}.
We have considered here ten of these evolutions, including the ones identified as
\emph{case\_1} and \emph{case\_3} in \citet{2016Icar...266..142B}. The simulations lasted
for 10 Myr. We recall that terrestrial planets have not been included
in this simulations.

\section{Results}

As expected from our previous study, the jumping Jupiter instability
disperse the hypothetical families through two different mechanisms:
(i) scattering of the asteroids due to close encounters with the fifth
giant planet before its ejection, and (ii) interaction of the asteroids
with sweeping secular and mean motion resonances with the planets.
The first mechanism is responsible for the dispersion in semi-major
axis, while the second causes the dispersion in eccentricity and inclination.
We found that all the jumping Jupiter evolutions considered here were 
able to scatter Vesta-like family members beyond 2.5~au. The scattering efficiency
depends on two factors. The first, and most important one, is the specific 
evolution of the fifth giant. In some evolutions, this planet reaches heliocentric distances as small as
1.5~au, thus strongly sweeping and scattering the inner part of the main belt.
In other evolutions, the fifth planet does not reach the inner belt and the scattering effect is
minimum. The second factor that influences the scattering efficiency is the particular 
initial configuration of the family.

\begin{figure*}
\begin{centering}
\includegraphics[width=1\textwidth]{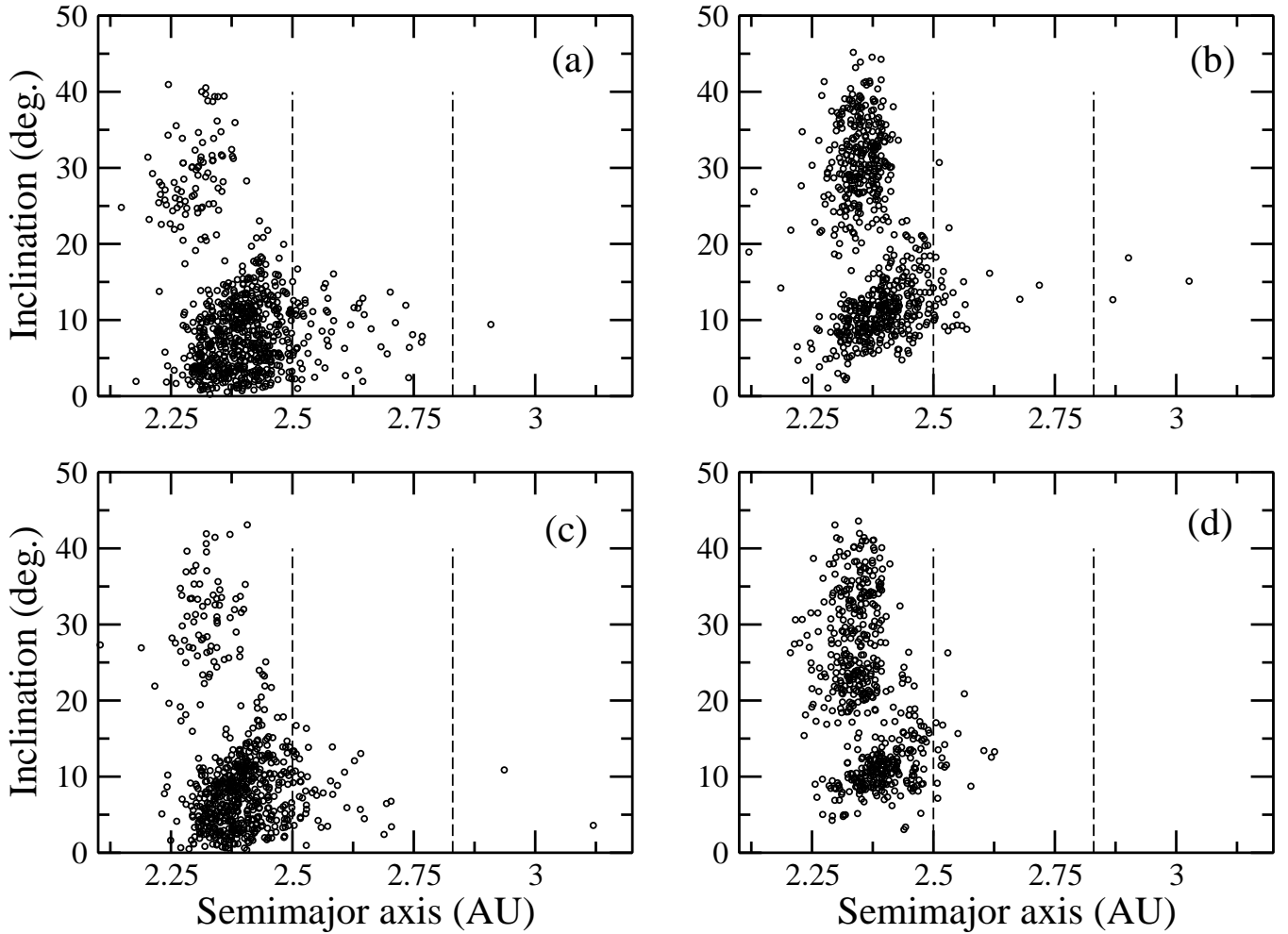}
\par\end{centering}
\caption{The final orbital distribution of four hypothetical Vesta-like families
in the inner asteroid belt, after evolving through a jumping Jupiter
instability (\emph{case\_1}). All the families are initially very compact, consistent
with a mean ejection velocity of 100~m~$\mathrm{s}^{-1}$. The different
panels correspond to the different orbital parameters of the parent body:
(a) $e=0.089$, $I=7.14^{\circ}$; (b) $e=0.089$, $I=15^{\circ}$;
(c) $e=0.2$, $I=7.14^{\circ}$; (d) $e=0.2$, $I=15^{\circ}$. In
all cases, the parent body had $a=2.36$~au, and each family had 1000
members. The vertical dashed lines mark the boundaries between the
inner, middle, and outer belt. The gap in the inner belt between $I\sim15^{\circ}$-$20^{\circ}$
is related to the $\nu_{6}$ secular resonance.}
\label{Brasil-fig3}
\end{figure*}

Figure \ref{Brasil-fig3} shows the typical results for a given 
jumping Jupiter evolution causing moderate scattering (\emph{case\_1}) and the four
initially compact families. These figures resemble the
situation observed in Fig. \ref{Brasil-fig1}, i.e. a concentration of
asteroids in the inner belt, and a few isolated bodies in the middle/outer
belt. The gap that is observed in the inner belt around $I\sim15^{\circ}$-$20^{\circ}$
is related to the $\nu_{6}$ secular resonance. In principle, the
group of orbits with $I>20^{\circ}$ is mostly in the region of terrestrial
planets crossing orbits and should be later eliminated by close encounters
in a few $10^{7}$ years if the terrestrial planets were included
in the simulations. As we will see later, the terrestrial planets will also be
responsible for the long term depletion of most of the orbits
that remain in the inner belt after the instability.

Our results should be analysed from a probabilistic point of view,
and we should not expect to precisely reproduce the current distribution
of the non-vestoid V-type asteroids. Nevertheless, we have verified
that, in general, the final inclinations of the dispersed family members
is correlated to the initial inclination of the parent body. Therefore,
the low inclination population of non-vestoids indicated in Fig.
\ref{Brasil-fig1} by red full circles could be pointing to
a hypothetical parent body with initially low inclination.

\begin{figure*}
\begin{centering}
\includegraphics[width=1\textwidth]{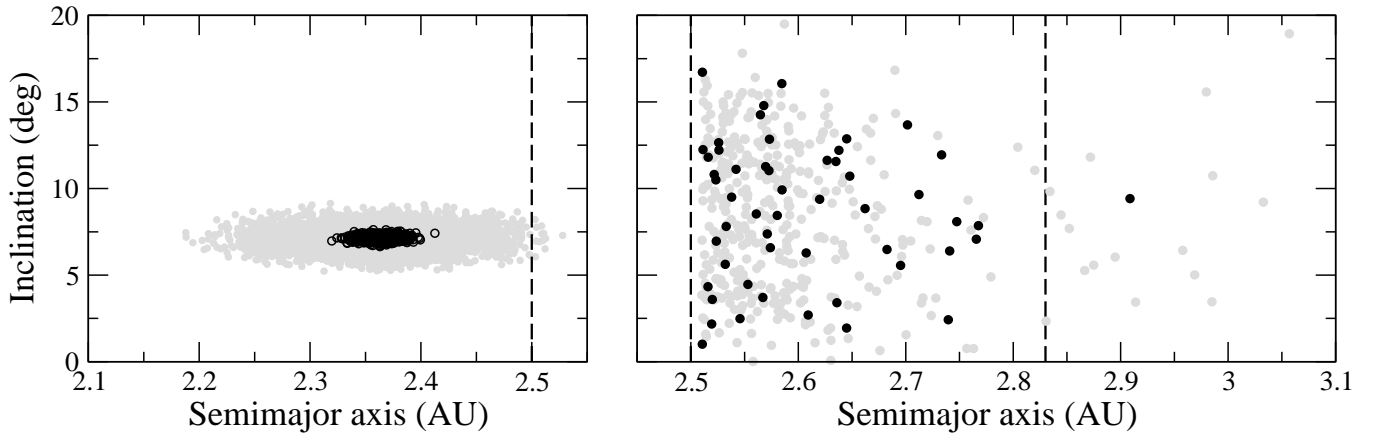}
\par\end{centering}
\caption{\emph{Left:} The initial orbital distribution of two synthetic families
in the inner belt generated using two different values of mean ejection
velocity. The more compact family, shown in black, had 1000 members
and corresponds to $\bar{v}_{\mathrm{ej}}=100$~m~$\mathrm{s}^{-1}$
(same case as panel (a) of Fig. \protect\ref{Brasil-fig3}). The more
extended family, shown in gray, had 5000 members and corresponds
to $\bar{v}_{\mathrm{ej}}=400$~m~$\mathrm{s}^{-1}$. \emph{Right:}
Final orbital distribution of the two families, for $a>2.5$~au, after
evolving through a jumping Jupiter instability (\emph{case\_1}).}
\label{Brasil-fig4}
\end{figure*}

We have also analysed the dependence on the mean ejection velocity
and the initial number $N$ of family members. In Fig. \ref{Brasil-fig4},
we show the two extreme cases: a very compact family with 1000 members
(in black), and a very extended family with 5000 members (in gray). The right
panel shows the final orbits of members scattered beyond 2.5~au. We
have verified that the amount of the scattered orbits roughly scales
linearly with $N$, and that the larger initial dispersion of the
family favours the scattering of orbits to longer distances. None
of our simulations scatters orbits beyond 3.2~au (the 2J:-1A mean motion 
resonance is located at 3.2~au), in agreement with
the fact that no confirmed V-type asteroid is currently known in that region
of the belt.

\begin{table}
\caption{Maximum implantation probabilities in three different target regions 
obtained for the eight initial configurations considered in this study.
Each configuration is identified
with the same letter used in Fig. \protect\ref{Brasil-fig3}, and the {*} symbol
refers to the very dispersed initial families with 5~000 members.
The target regions correspond to specific locations where non-vestoids are found
(see text).}

\begin{centering}
\begin{tabular}{cccc}
\hline 
\multirow{2}{*}{Configuration} & \multicolumn{3}{c}{$p_{\mathrm{imp}}$} \tabularnewline
\cline{2-4}
       & Middle belt & Outer belt & Low-$I$ inner belt \tabularnewline
\hline 
(a)    & 10.0\%  & 0.8\%  & 40.9\%  \tabularnewline
(a){*} & 12.3\%  & 0.9\%  & 31.3\%  \tabularnewline
(b)    &  8.8\%  & 1.4\%  &  3.6\%  \tabularnewline
(b){*} & 10.1\%  & 0.9\%  &  3.7\%  \tabularnewline
(c)    & 11.0\%  & 0.6\%  & 21.8\%  \tabularnewline
(c){*} & 11.9\%  & 1.1\%  & 17.9\% \tabularnewline
(d)    &  8.8\%  & 0.9\%  &  1.2\%  \tabularnewline
(d){*} &  9.9\%  & 1.5\%  &  1.3\% \tabularnewline
\hline 
\end{tabular}
\par\end{centering}
\label{table1} 
\end{table}

In Table \ref{table1}, we summarise the implantation probabilities obtained 
for each of the eight family configurations considered in our study. These
probabilities represent the maximum values computed over the whole set of 
jumping Jupiter evolutions analysed here (which means that, for some 
evolutions, a given probability may actually be zero). We
tested three target regions where orbits can be implanted. These
correspond to the locations where 
the non-vestoids with no determined dynamical link
to (4)~Vesta or to the Vesta family are mostly found, i.e. the middle main belt,
the outer main belt, and the low inclination inner main belt.

The implantation probabilities in the outer belt are an order of magnitude smaller that those
in the middle belt. A family initially dispersed may implant up to 2 times more
objects in the outer belt than an initially compact one. In the middle belt, however,
the implanted probabilities shows no significant dependence on the family compactness.
A different situation is observed in the low inclination inner belt, where the
implantation probabilities are clearly correlated to the initial inclination
of the family.

To determine how reliable are these probabilities,
we estimate the number of asteroids, $N_{\mathrm{src}}$, that
should be in the original hypothetical family to reproduce the currently
observed populations in the three target regions. This is given by
the formula:
\begin{equation}
N_{\mathrm{src}}=\frac{N_{\mathrm{obs}}}{f_{\mathrm{obs}}\times p_{\mathrm{imp}}\times f_{\mathrm{surv}}}
\label{eq1}
\end{equation}
where $N_{\mathrm{obs}}$ is the number of V-type asteroids observed
in the target region up to a given absolute magnitude $H$, $f_{\mathrm{obs}}$
is the fraction of all the existing V-type asteroids that have been
observed up to the same $H$, $p_{\mathrm{imp}}$ is the implantation
probability in the target region given in Table \ref{table1}, and
$f_{\mathrm{surv}}$ is the fraction of implanted orbits after the instability that
would be able to survive until present days.

In the spectroscopic sample, $N_{\mathrm{obs}}$ is well determined
but it is not statistically significant, especially in the middle/outer
belt. Moreover, the spectroscopic sample is strongly affected by observational
biases related to the different surveys, which makes the determination
of $f_{\mathrm{obs}}$ quite difficult. On the other hand, the SDSS or the MOVIS
samples of V-type candidates provide more significant values of $N_{\mathrm{obs}}$
and allows for a more reliable determination of $f_{\mathrm{obs}}$.
To account for the fact that we are dealing with photometric candidates,
we replace $N_{\mathrm{obs}}=N_{\mathrm{cand}}\times f_{\mathrm{conf}}$,
where $N_{\mathrm{cand}}$ is the actual number of candidates in the
sample and $f_{\mathrm{conf}}$ is the fraction of these candidates
that are expected to be confirmed as V-type by spectroscopic observations.
 Based on previous results by several authors (e.g. \citealp{2013A&A...552A..85J}, 
and other references already cited here), we estimate that $f_{\mathrm{conf}}\sim0.8$.

To estimate $f_{\mathrm{obs}}$ for the SDSS or the MOVIS sample, we adopted two approaches.
The first one consists into divide the total number of main belt asteroids observed in either 
the SDSS or MOVIS sample 
by the total number of known main belt asteroids, considering a cut off in absolute magnitude
$H\leq15$. According to \citet{2015aste.book..795J}, this is the limiting magnitude 
for which the main belt sample is 
complete, and there are $\sim 125\,000$ main belt asteroids with $H\leq15$.
From the SDSS and MOVIS samples, we get 20\,600 (\citealp{2010A&A...510A..43C}) and
17\,200 (\citealp{2016A&A...591A.115P}) main belt asteroids with $H\leq15$, respectively.
This gives a value of $f_{\mathrm{obs}}$ between 0.14 and 0.16.
The second approach consists into divide the number of SDSS or MOVIS 
V-type candidates found in the Vesta family by the total number of known members of the family, also with
a cut off $H\leq15$. According to \citet{2015PDSS..234.....N}, there are 2\,408 members of the
Vesta dynamical family with $H\leq15$, but according to \citet{Licandro2017} only $\sim85$\%
of these would be actual V-type asteroids, which means $\sim2\,050$ members. 
From the SDSS and MOVIS samples, we get 350 (\citealp{2010A&A...510A..43C}) and
233 (\citealp{Licandro2017}) V-type candidates with $H\leq15$, respectively. This
implies a value of $f_{\mathrm{obs}}$ between 0.11 and 0.17, in good agreement
with the previous estimate. Here, we will assume $f_{\mathrm{obs}}=0.15$.

Finally, to estimate $f_{\mathrm{surv}}$, we use the simulations by \citet{2015AJ....150..186R}
and \citet{Nesvorny2017}. These authors simulated the evolution of the primordial asteroid
belt over the age of the solar system considering four different phases: (i) the evolution before
the jumping Jupiter instability during which the giant planets are in a compact orbital
configuration and do no migrate, (ii) the evolution during the instability, (iii) the evolution 
after the instability during which the giant planets continue to migrate
smoothly until the planetesimals disk is totally dispersed and the
planets reach their present orbits (this phase of residual migration
may last between 100 and 300 Myr), and (iv) the evolution after migration ceased
and until the present day (typically lasting $\sim4$~Gyr). 
In the case of \citet{Nesvorny2017}, the simulations include the 
gravitational perturbation of the terrestrial planets. This latter effect has proven to be relevant
for the long term depletion of the inner main belt after the jumping Jupiter instability,
especially during phase (iv).
Actually, the inner belt is depleted by a factor of $\sim2$ compared to the simulations where
the terrestrial planets are not considered. A similar behaviour is not observed in the middle
and outer belt, where the long term influence of the terrestrial planets is not so relevant.
Based on these results, we estimate that, after the instability, the fractions of orbits that survive
until today in the inner, middle and outer main belt are $f_{\mathrm{surv}}=0.14$, 0,66 and 0.59,
respectively.

The resulting estimates of $N_{\mathrm{src}}$ for each target region are given in Table \ref{table2}.
These numbers can be compared to the $\sim2400$ members of the current Vesta
family with $H\leq15$ to conclude that a Vesta-like paleo-family
could have been the source of many current non-vestoids in the three
regions, scattered during the jumping Jupiter instability. 
The only clear exception is the case of a high-$I$ paleo-family which produces
totally unrealistic estimates for the low-$I$ inner belt non vestoids. This 
strengths the conclusion that the very low inclination non-vestoids in the inner belt
should be related to a parent body with a Vesta-like initial inclination.
In the remaining cases, the upper limit estimates of $N_{\mathrm{src}}$ obtained for the three regions
are not critical since the main source of uncertainty
in the above calculations is the maximum $p_{\mathrm{imp}}$, which
depends on the initial configuration of the family. These upper limit estimates 
may also be reduced by assuming
that some of the non-vestoids in the three regions were not scattered
by the jumping Jupiter instability, but have different origins.

\begin{table*}
\caption{The estimated number $N_{\mathrm{src}}$ of asteroids with $H\leq15$
that would be in the initial hypothetical family to reproduce the amount of observed SDSS or MOVIS non-vestoids
$N_{\mathrm{cand}}$ in the different regions. For the low inclination
inner belt, we give two estimates depending on the two extreme values
of $p_{\mathrm{imp}}$ obtained for the low/high initial inclination of the family.
The last column gives the estimated diameter range of the crater that
should be excavated to generate the hypothetical family, assuming a
crater depth $h=1$~km.}

\begin{centering}
\begin{tabular}{ccccc}
\hline 
\multirow{2}{*}{Region} & $N_{\mathrm{cand}}$  & $N_{\mathrm{cand}}$  & \multirow{2}{*}{$N_{\mathrm{src}}$}   &  $d_{\mathrm{crat}}$    \tabularnewline
                                   & (SDSS)                       & (MOVIS)                      &    &  (km) \tabularnewline
\hline 
Middle belt   & 23  &  13   &    850--2\,100  &  115--180   \tabularnewline
Outer belt\smallskip{}     & 12  &    6   &  3\,600--18\,000  &  240--540 \tabularnewline
\multirow{2}{*}{Low-$I$ inner belt} & \multirow{2}{*}{60} & \multirow{2}{*}{16} & 1\,500--12\,700   &  160--450  \tabularnewline
 &  &  & 16\,500--190\,000  &  --  \tabularnewline
\hline 
\end{tabular}
\par\end{centering}
\label{table2} 
\end{table*}

Using the values of $N_{\mathrm{src}}$, we may estimate the 
total mass that should be excavated on the parent body. For this
calculation, we assume that the cumulative size distribution of the fragments follows
a power law $N(>D)=A\,D^{-\alpha}$ with exponent $\alpha =5$. 
This is similar to the SFD of the 
current Vesta family in the range $2.1\leq D\leq 8.0$~km (which corresponds to
$12.1\leq H\leq 15.0$ for albedo 0.4). The value of $A$ is calibrated by imposing
$N(>D_0)=N_{\mathrm{src}}$ for $D_0=2.1$~km ($H=15$), and the diameter of the 
largest fragment
is then given by $D_{\mathrm{max}}=A^{1/\alpha}$. The total mass excavated
can be computed as:
\begin{equation}
M(>D)=-\frac{\pi}{6}\rho\alpha A\int_{D_{\mathrm{max}}}^{D}D'^{-3(\alpha+1)}dD'
\end{equation}
where $\rho=3.0$~g~cm$^{-3}$ is the typical density of basalt.
Assuming that $N_{\mathrm{src}}\gg 1$ and $D_0\geq D$, the above expression
reduces to:
\begin{equation}
M(>D)\simeq\frac{\pi}{6}\rho\frac{\alpha}{3\alpha+2}\frac{D_{0}^{\alpha}N_{\mathrm{src}}}{D^{3\alpha+2}}
\label{MD}
\end{equation}
On the other hand, the excavated mass is roughly related to the 
size $d$ and depth $h$ of the corresponding 
crater (in km) through the formula: 
\begin{equation}
M\simeq \frac{\pi}{6}\rho h\left(\frac{3}{4}d^{2}+h^{2}\right),
\label{MV}
\end{equation}
and equating (\ref{MD}) and (\ref{MV}) we obtain:
\begin{equation}
d\simeq \left(\frac{4\alpha}{9\alpha+6}
\frac{D_{0}^{\alpha}N_{\mathrm{src}}}{D^{3\alpha+2}h}
-\frac{4}{3}h^{2}\right)^{1/2}
\end{equation}
The last column in Table \ref{table2} summarises the estimated crater diameters 
assuming $D=1$~km and a crater depth $h=1$~km. 
If the depth is reduced to 0.5~km, the diameters increase by a factor
$\sqrt{2}$. Typical values of $\alpha$ may vary between 3 and 7, which 
implies that the diameters may be reduced/increased by a factor of up to 2. 
These diameters are smaller than or comparable to the diameters of the
largest basins on the surface of Vesta.

We have also determined whether the final orbital distribution of
implanted orbits reproduces the current distribution of the non-vestoids.
Following \citet{2014MNRAS.439.3168C} and \citet{2014MNRAS.444.2985H},
we have divided the middle and outer belt in seven subregions that
according to these authors are dynamically isolated in the long term,
i.e. asteroids have a low probability of migrating from one region
to another even over Gyr timescales. We have verified that our simulations
are able to implant orbits in all the seven subregions depending on 
the initial configuration of the family as well as on the specific jumping
Jupiter evolution.
Table \ref{table3} shows the fractions of implanted orbits beyond 2.5~au that are implanted
into each of the seven subregions. The fractions have been averaged over the
different initial configurations of the family, separating the low inclination
configurations (e.g. panels (a),(c) in Fig. \ref{Brasil-fig3}) 
from the high inclination ones (e.g. panels (b),(d)
in Fig. \ref{Brasil-fig3}). The maximum and minimum values of each fraction reflects
the effect of the different jumping Jupiter evolutions.

Several features can be addressed from these results. For example, the largest fraction
of implanted bodies is concentrated in the interval 2.5-2.7~au. Some jumping Jupiter 
evolutions are not able to implant bodies beyond 2.7~au. Once again, there is a correlation
between the initial inclination of the family and the final inclination of the
implanted orbits, especially in the middle belt. The last rows in Table \ref{table3}
show the fractions of the SDSS and MOVIS V-type candidates identified in the same seven
subregions. In some regions, we observe a pretty good match, but in other regions
the implanted fractions are either underestimated or overestimated by a factor that
typically ranges between 2 and 5. In particular, our simulations do not seem to produce 
enough V-type asteroids in the outer main belt in what is called 
the Eos region (sixth data column). We must bear in mind, however, that the
fractions reported in the last two rows may be affected by poor statistics.

\begin{table*}
\caption{Fractions of the implanted orbits that are implanted into each of the
seven subregions defined in the middle and outer belt. Simulations
are identified using the same convention as in Table \protect\ref{table2}.
The last two rows give the same fractions from the SDSS and MOVIS samples of V-type
candidates.}

\begin{centering}
\begin{tabular}{ccccccccc}
\hline 
\multirow{2}{*}{Configurations} & \multicolumn{2}{c}{$2.5\leq a<2.71$~au} &  & \multicolumn{2}{c}{$2.71\leq a<2.83$~au} & \multirow{2}{*}{$2.83\leq a<2.97$~au} & \multicolumn{2}{c}{$2.97\leq a<3.28$~au}\tabularnewline
\cline{2-3} \cline{5-6} \cline{8-9} 
 & $I\leq6.9^{\circ}$ & $I>6.9^{\circ}$ &  & $I\leq6.9^{\circ}$ & $I>6.9^{\circ}$ &   & $I\leq14.9^{\circ}$ & $I>14.9^{\circ}$\tabularnewline
\hline 
(a), (a){*}, (c), (c){*} & 28.6\%--44.0\%  & 54.8\%--56.9\%  &  & 0\%--3.1\%  & 0\%--6.8\%  &  0\%--4.5\%  &  0.6\%--1.4\%  & 0\%--0.3\% \tabularnewline
(b), (b){*}, (d), (d){*} & 0\%--5.9\%  & 71.4\%--100\%  &  & 0\%--0.8\%  & 0\%--9.9\%  &  0\%--7.6\%   & 0\%--1.9\%  & 0\%--2.5\% \tabularnewline
SDSS sample  & 32\%  & 26\%  &  & 8\%  & 14\%  &  6\%   & 13\%  & 1\% \tabularnewline
MOVIS sample  & 16\%  & 37\%  &  & 5\%  & 11\%  &  0\%   & 26\%  & 5\% \tabularnewline
\hline 
\end{tabular}
\par\end{centering}
\label{table3}
\end{table*}

A final issue refers to the amount of fragments from the hypothetical family
that would survive until present days in the inner belt (excluding the low-$I$ region). 
These survivors are expected to be mixed today with the V-type asteroids created at
later times by the Rheasilvia and Veneneia impact events. The amount of these bodies can be estimated as 
\begin{equation}
N_{\mathrm{src}}\times (1-f_{\mathrm{deplet}})\times f_{\mathrm{surv}}
\label{eqdep}
\end{equation}
where $f_{\mathrm{deplet}}$ is the fraction of orbits that are depleted from
the inner belt during the jumping Jupiter instability, and $N_{\mathrm{src}}$ and
$f_{\mathrm{surv}}$ are the parameters involved in Eq. (\ref{eq1}). As previously
shown, $N_{\mathrm{src}}$ ranges between 850 and 18\,000, while 
$f_{\mathrm{surv}}=0.14$ for the inner belt. 
The values of $f_{\mathrm{deplet}}$
can be determined from the simulations performed in this work, and we found values ranging between
0.4 and 0.9, mostly depending on the specific jumping Jupiter evolution. It is worth noting
that the largest depletion fractions are provided precisely by the evolutions that are more efficient
in implanting orbits into the middle/outer belt, so the large values of $f_{\mathrm{deplet}}$ 
should be preferred in Eq. (\ref{eqdep}) over the smallest ones. Therefore, using a compromise value of $f_{\mathrm{deplet}}=0.7$, we
obtain that the amount of surviving fragments would be something between 36 and 750, or even smaller. 
An additional depletion factor that may contribute to reduce even more these numbers can be
provided by the Yarkovsky effect, which is relevant for V-type asteroids with $H\gtrsim12$ over Gyr
timescales. Multiplying these number of survivors by the fraction of asteroids with known taxonomic
classification, $f_{\mathrm{obs}}=0.15$, we found that the number of currently observed V-type asteroids
that may be related to an hypothetical Vesta-like paleo-family is between 5 and 110. This
represents, for example, less than 13\% of the non-vestoid SDSS candidates in the inner belt 
(excluding the non-vestoids in the low-$I$ region).

\section{Conclusions}

In this work, we have analysed the effect of the jumping Jupiter instability
on a hypothetical primitive family of V-type asteroids in the inner
belt. Our working hypothesis was that such family formed from a craterization
event on the surface of (4)~Vesta during the epoch of planetary migration
of the giant planets, around or more than 4 Gyr ago. This hypothesis
is supported by data from the Dawn mission and simulations of the
collisional history of (4)~Vesta (\citealp{2016Icar..271..170P}).
Our main conclusions can be summarised as follows:
\begin{itemize}
\item The jumping Jupiter instability is able to scatter some of the V-type
asteroids of this hypothetical family into regions of the Main Belt
that otherwise could not be reached by long term dynamical evolution.
This mechanism may explain, at least partially, the population of non-vestoids observed
in the middle and outer belt.
\item The instability can also explain the population of non-vestoids
with very low inclinations in the inner belt ($I\lesssim 5^{\circ}$), provided that the parent body had
a Vesta-like inclination ($\sim 7^{\circ}$).
\item We estimate that the diameter of the crater that has to be excavated to originate the 
hypothetical primitive family ranges between 100 and 500~km. According to \citet{2016Icar..271..170P},
traces of such a basin might have been erased by the subsequent collisional history of (4)~Vesta.
\item We estimate that $\sim 10$\% or less of the currently observed V-type asteroids
in the inner belt may be relics of this primitive family.
\item Although the jumping Jupiter instability can implant inner belt asteroids
in Magnya-like orbits, this is not enough to explain the origin of
(1459)~Magnya. The typical implantation probabilities imply that
in order to obtain (1459)~Magnya there should have been several tens
of V-type asteroids with $D\sim20$~km in the original source. This
hypothesis is incompatible with a craterization event and is not supported
by observations, since no other V-type asteroid in this size range
has been detected so far.
\item Our model suggests that many non-vestoids, including those in the
middle/outer belt \textendash except of (1459)~Magnya\textendash ,
should be genetically related to (4)~Vesta but not to the current
Vesta family. Observational evidence from spectroscopy and polarimetry,
in principle, is not incompatible with this idea.
\item (1459)~Magnya continues to be a unique piece of basaltic material
whose origin is still an open question.
\end{itemize}

\section*{Acknowledgments}

We wish to thank the helpful comments and criticism of the referees.
P.I.O.B. and F.R. acknowledge support from the Brazilian Council of
Research (CNPq). D.N. is supported by NASA's Emerging Worlds program
and Brazil's Science without Borders program. V.C. is supported by
the S\~{a}o Paulo State Science Foundation (FAPESP).



\bsp

\label{lastpage}
\end{document}